\documentclass[review]{elsarticle}

\usepackage{lineno,hyperref}
\usepackage{amsfonts}
\usepackage{graphicx}
\usepackage{balance}
\usepackage{comment}
\usepackage{amsmath}
\usepackage{color}
\usepackage{url}
\usepackage{subfigure}
\usepackage[ruled]{algorithm}
\usepackage{algorithmic}
\usepackage{empheq}
\usepackage{fancyhdr}

\journal{Journal of \LaTeX\ Templates}

\begin{document}

\begin{frontmatter}

\title{Distributed Least-Squares Iterative Methods in Networks: A Survey}

\author[GSU]{Lei Shi}
\ead[GSU]{lshi1@student.gsu.edu}
\author[GGC]{Liang Zhao}
\ead[GGC]{lzhao2@ggc.edu}
\author[UGA]{Wen-Zhan Song}
\ead[UGA]{wsong@uga.edu}
\author[GSU]{Goutham Kamath}
\ead[GSU]{gkamath1@student.gsu.edu}
\author[ZJUT]{Yuan Wu}
\ead[ZJUT]{iewuy@zjut.edu.cn}
\author[HKPU]{Xuefeng Liu}
\ead[HKPU]{csxfliu@comp.polyu.edu.hk}
\address[GGC]{Department of Computer Science, Georgia Gwinnett College}
\address[GSU]{Department of Computer Science, Georgia State University}
\address[UGA]{College of Engineering, University of Georgia}
\address[ZJUT]{College of Information Engineering, Zhejiang University of Technology}
\address[HKPU]{Department of Computing, The Hong Kong Polytechnic University}






\begin{abstract}
Many science and engineering applications involve solving a linear least-squares system formed from some field measurements. In the distributed cyber-physical systems (CPS), often each sensor node used for measurement only knows partial independent rows of the least-squares system. To compute the least-squares solution they need to gather all these measurement at a centralized location and then compute the solution. These data collection and computation are inefficient because of bandwidth and time constraints and sometimes are infeasible because of data privacy concerns. Thus distributed computations are strongly preferred or demanded in many of the real world applications e.g.: smart-grid, target tracking etc. To compute least squares for the large sparse system of linear equation iterative methods are natural candidates and there are a lot of studies regarding this, however, most of them are related to the efficiency of centralized/parallel computations while and only a few are explicitly about distributed computation or have the potential to apply in distributed networks. This paper surveys the representative iterative methods from several research communities. Some of them were not originally designed for this need, so we slightly modified them to suit our requirement and maintain the consistency. In this survey, we sketch the skeleton of the algorithm  first and then analyze its time-to-completion and communication cost. To our best knowledge, this is the first survey of distributed least-squares in distributed networks. \footnote{Our research is partially supported by NSF-CNS-1066391, NSF-CNS-0914371, NSF-CPS-1135814 and NSF-CDI-1125165.}

\end{abstract}


\end{frontmatter}


\section{Introduction}
\label{sec:intro}
Many physical phenomena can be described by partial differential equation~\cite{B-NMLSP1996} which when further discretized forms large sparse system of linear equations. Many important problems, such as state estimation, target tracking and tomography inversion, are often formulated as a large scale linear system based on some field measurements. Those field measurements may contain errors, thus extra amount of measurements are often sampled to form an over-determined linear system:
 
\begin{equation}
  \label{eqn:eqnsys}
  Ax\approx b
\end{equation}
where $A \in \mathbb{R}^{m\times n} (m\geq n)$ and $x \in \mathbb{R}^{n}$ and $b \in \mathbb{R}^{m}$. 
These extra information smoothed out the errors but produced an overdetermined system which usually had no exact solution. The method of least-squares is a common approach to obtain the solution to the above problem and can be defined as
\begin{equation}
  \label{eqn:ls}
  \min_{x}\|Ax-b\|_2
\end{equation}

The coefficient in $A$ is often modeled from the data obtain from sensors used for observing the physical phenomena eg: cyber physical system. Each sensor or node observes partial phenomena due to the spatial and temporal restriction and thus only form partial rows of the least-squares systems. The large-scale cyber-physical systems are often built on a mesh network, which could be a wired or wireless or wired-wireless hybrid multi-hop network. For instance the problem from target tracking, seismic tomography and the smart grid state estimation problem all have an inherently distributed system of linear equations. However, the least squares method used currently to obtain the solution to these problem assume a centralized setup where partial row information from all the nodes are collected in a server and then solved using centralized least-square algorithm.

In many of those cyber-physical systems, the distributed computation in mesh networks is strongly demanded or preferred over the centralized computation approach, due to the following reasons (but not limited to): (1) In some applications such as imaging seismic tomography with the aid of mesh network, the real-time data retrieval from a large-scale seismic mesh network into a central server is virtually impossible due to the sheer amount of data and resource limitations. While the distributed computation may process data inside the network in real-time to reduce the bandwidth demand as well as distribute the communication and computation load to each node in the network. (2) The mesh network can be disruptive in real world, the data collection and centralized computation may suffer from the node failure or link disruption. It becomes a bottleneck especially if the node failure or link disruption happens near to the sink node which leads to loss of high volume of raw data. With the distributed computation remaining nodes in the network can still finish the computation and get the approximated results. (3) In smart grid state estimation, the data collection for centralized computation is even infeasible due to the data privacy concerns or inter-agency policy constraints. (4) In some applications that needs real-time control, the distributed computation will also have advantage over centralized scheme, since some decisions can be made locally in real-time. The current state of the art computational device such as smart-phones etc enables us to perform in-network computing and carry out distributed computation over a mesh network.

Iterative methods are natural candidates when it comes for large sparse system and especially for distributed computation of least-squares. Although there are a lot of studies on iterative least-squares, most of them are concerned with the efficiency of centralized/parallel computations, and only a few are explicitly about distributed computation or have the potential to applied on mesh networks. In mesh network, since the computers may need to communicate with each other through message passing over a multi-hop network, the key challenges  are not only speeding up the computation but also reducing the communication cost. Often more attention shall be paid to communication than the computation cost, especially when solving a big problem in a large scale mesh network. Thus, in this paper we select and survey the representative iterative methods from several research communities that have the potential to be used in solving least squares problem over mesh network. Here, a skeleton sketch of each algorithm is provided and later we analyze its time-to-completion and communication cost and provide the comparison. Some of the algorithms presented here were not originally designed for to meet our requirement, so we slightly modify them to maintain consistency. To our best knowledge, this is the first attempt to survey distributed algorithms from different domains that is suitable to perform least-squares over mesh networks.

The rest of the paper is organized as follows. In section \ref{sec:netmod}, we present the network model and the evaluation criteria for comparison. Then in section \ref{sec:DisAlg}, we describe the state of art and each surveyed algorithm in details, analyze and compare their communication costs and time-to-completion. Finally, we conclude the paper in section \ref{sec:conc}.

\section{Model and Assumption}
\label{sec:netmod}

Consider a wired and/or wireless mesh network with $N$ nodes $v_1,\ldots,v_N$ which form connected graph and can be reached through multi-hop message relays. Without loss of generality, we can assume that the diameter of the network is $\log N$ (i.e., any message can be sent from one node to another through at most $\log N$ hops). Also, let us assume that each node has a single radio and the link between the neighboring node has a unit bandwidth. Therefore, the communication delay of one unit data delivery between direct neighbors (either through a unicast to one direct neighbor or multicast/broadcast to all direct neighbors) would be one unit time. Notice that, here we assume the link layer supports broadcast which is often true in many mesh networks. If the link layer only supports unicast, the analysis can be similarly done by considering a one-hop broadcast as multiple unicast and it is omitted here due to page limit. We know that the link layer communication may take more than one unit time for one unit data due to network interference and media contentions. Thus, we classify the communication pattern in mesh network into three categories, unicast (one-hop or multi-hop), one-hop broadcast (local broadcast to all neighbors) and network flooding (broadcast to all the nodes in network). For simplicity and convenience, in the rest of the paper we use term \emph{broadcast} for local broadcast to one-hop neighbors and \emph{flood} for network flooding. We use the aforementioned assumption on communication cost and delay for the quest of the fundamental limit of each surveyed algorithm in an ideal mesh network.

For comparison and evaluation, the following performance criteria is considered:
\begin{itemize}
    \item Communication cost: To solve a least-squares problem of large size, the communication cost has a big influence on the algorithm performance. Here we refer the communication cost as the cost involved in the messages exchanged in the mesh network during a single iteration of the iterative methods. Since iterative methods typically converge after many iterations, the communication cost of the iterative methods depends on both the cost in one iteration and the iteration number.
    \item Time-to-completion: The time taken for the network to finish one iteration in the iterative method is referred as time-to-completion in this paper. Notice that, it is different from the computational time complexity: the time-to-completion shall include the consideration of the message size and number of hops the packet traversed. Also, in this work, we focus on the analysis of communication delays for time-to-completion while ignoring the computation time in each node. With the Moores law, the computation capability are increasing faster than communication capacity of transceivers and typically transceivers consume more power and also use dominate the computational time. Therefore, these two criteria become important in-order to compare various types of distributed least-squares methods in our survey paper. 
\end{itemize}

\begin{figure}[htb!]
\centering
\includegraphics[width=3.4in]{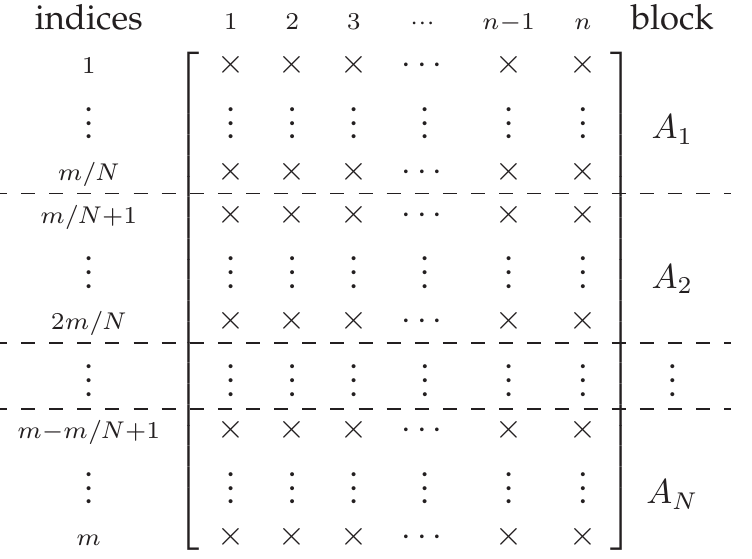}
\caption{Row partition of matrix $A$}
\label{fig:rowpart}
\end{figure}

Least-squares problem (see equation~\ref{eqn:ls}) formed over the mesh network are inherently distributed i.e. each node $v_u$ only knows part of $A$ and $b$. We assume that each node in the network holds $m_u=m/N$ consecutive rows of matrix $A \in \mathbb{R}^{m \times n}$ and the corresponding part of vector $b$. For example in Figure~\ref{fig:rowpart} block $A_{1}$ indicates the first $m/N$ rows of matrix $A$ which is assigned to node $v_1$ along with the right hand side vector $b = \{b_1,\ldots,b_{m/N}\}$. Note that the algorithm surveyed in this paper do not require that matrix $A$ and $b$ be equally partitioned over the network. Here the assumption of equal partition is for the simplicity of presentation and analysis and the new distributed equation will take the form,

\begin{equation}
\label{eqn:Axb}
Ax = b
\end{equation}

where,
\begin{equation*}
A = \begin{pmatrix} 
							A_{1} \\
							A_{2} \\
							\vdots \\
							A_{N}
							\end{pmatrix} ; b = \begin{pmatrix} 
							b_{1} \\
							b_{2} \\
							\vdots \\
							b_{N}
							\end{pmatrix} ; A_{u} \in \mathbb{R}^{m_{u} \times n};b_{u} \in \mathbb{R}^{m_{u}}
\end{equation*}

The least squares problem takes the form $\min_{x}\|Ax-b\|_2$ and since there is no central coordinator which has entire $A$ and $b$ the computation of optimum $x$ has to be done distributedly. As mentioned above communication cost becomes crucial for distributed solution over sensor network and the goal of this paper is to survey various distributed least squares algorithm originating from different domains. We also try to compare different algorithms under similar criteria as mentioned above so that it provides the reader some basic differences between them and also help them to choose the type of algorithm suitable for their application. 

The notations used in this paper are described in Table \ref{tbl:symbols}.

\begin{center}
\begin{table}[htbp]
\caption{List of notations used in this paper}
\label{tbl:symbols}
\centering
 \begin{tabular}{|l|l|}
 \hline
 $A, E, L, U, Q, R\ldots$ & matrices\\
  \hline
 $x, y, a, b, r\ldots$ & vectors\\
  \hline
 $\alpha, \beta, \delta, \lambda, \gamma\ldots$ & scalars\\
  \hline
 $m, n$ & rows and columns of matrices\\
  \hline
 $\mathbb{R}$ & real space\\
  \hline
 $N$ & network size\\
  \hline
 $D_{avg}, D_{max}$ & the average and maximum node degree in the network\\
 \hline
 $u, v$ & nodes in the network\\
 \hline
 $k$ & iteration number of iterative methods \\
 \hline
\end{tabular}
\end{table}
\end{center}

\section{Survey and Analysis}
\label{sec:DisAlg}
The methods to solve the linear least-squares problem are typically classified into two categories, direct methods and iterative methods. Direct methods are based on the factorization of the coefficient matrix $A$ into easily invertible matrices whereas iterative methods solve the system by generating a sequence of improving approximate solutions for the problem. Until recently direct methods were often preferred over iterative methods~\cite{SV-JCAM2000} due to their robustness and predictable behaviors (one can estimate the amount of resources required by direct solvers in terms of time and storage)~\cite{DER-CP1989, B-JCP2002, S-SIAM2003}. However a number of iterative methods were discovered which required fewer memory and started to approach the solution quality of direct solvers~\cite{S-SIAM2003}. The size of the least squares problem arising from real world three-dimension problem. models could be significantly large comprising hundreds of millions of equations as well as the unknowns. Despite such a huge dimension typically the matrices arising will be sparse and can be easily stored. Now given the dimension and sparsity property of the matrix, iterative methods become almost mandatory for solving them~\cite{GG-JSC2005}. Also, iterative methods are gaining ground because they are easier to implement efficiently on high-performance computers than direct methods~\cite{S-SIAM2003}.

To achieve high performance in computation, researchers have studied both parallel and distributed iterative methods to solve large linear systems/linear least-squares problems~\cite{HNP-SIAMR1991, B-THESIS2006}. The researches in parallel computing of large linear systems involve both shared and distributed memory architecture. Traditionally, the parallel computing is distinguished from distributed computing with memory architecture. In parallel computing, all computers may have access to a shared memory whereas in distributed networks, each computer has its own private memory (distributed memory) and information is exchanged by passing messages between the computers. Typically these message exchanges involve dedicated bus or high bandwidth communication channel which are relatively easier when compared to communication over mesh network. So, in this paper, we only present those distributed iterative methods or the parallel iterative methods that can be potentially distributed over a mesh network. For the parallel iterative algorithm candidates, we re-describe the original algorithm in the context of distributed computing based on the mesh network model discussed in previous section. The following algorithms are discussed and analyzed in details in this paper,

\begin{itemize}
 \item{\bf D-MS} Distributed Multisplitting method in section~\ref{subsec:Multisplitting}.
 \item{\bf D-MCGLS} Distributed Modified Conjugate Gradient Least-Squares method in section~\ref{subsec:MCGLS}.
 \item{\bf D-LMS} Distributed Least Mean Squares method in section~\ref{subsec:DLMS}.
 \item{\bf D-RLS} Distributed Recursive Least-Squares method in section~\ref{subsec:DRLS}.
\end{itemize}

Table~\ref{tab:commcost} gives a summary of the analysis of the communication cost and time-to-completion of the selected algorithms running in distributed network. The details about the algorithm description and analysis are shown in section \ref{sec:DisAlg}. Considering the least-squares problem in equation (\ref{eqn:ls}) where $A \in \mathbb{R}^{m\times n} (m\geq n)$, $x \in \mathbb{R}^{n}$ and $b \in \mathbb{R}^{m}$. Suppose that the iterative algorithm converges within $k$ iterations in the network, $D_{avg}$ and $D_{max}$ denote the average and maximum node degree of the network respectively. The algorithms discussed in this paper have been proved to be convergent, but the iteration number highly depends on the matrix condition number, these algorithms may need hundreds to thousands of iterations to converge over a network with hundreds of nodes for a large system. Besides, some algorithms either requires flooding communication in the network per iteration or a Hamiltonian path in the network to perform the computation node by node.

\begin{table*}[htb!]
\centering
\caption{Communication Cost and Time-to-completion Analysis}
\label{tab:commcost}
\begin{tabular}{|l|l|l|l|l|}
  \hline
  & Algorithm & Section & Communication Cost & Time-to-completion\\
  \hline
  1 & D-MS & \ref{subsec:Multisplitting} & $kmN^2$ & $km(N-1)$\\
  2 & D-MCGLS & \ref{subsec:MCGLS} & $(k+1)(m+N)N+k(n+N)N$ & $k(m+n+2)(N-1)$\\
  3 & D-LMS & \ref{subsec:DLMS} & $kN(D_{avg}+1)$ & $2kD_{max}$\\
  4 & D-RLS & \ref{subsec:DRLS} & $(n+n^2)(N-1)$ & $(n+n^2)(N-1)$\\
  \hline
\end{tabular}
\begin{center}
$N$ is the network size, $m\times n (m\geq n)$ is dimensions of matrix $A$ and $k$ is the number of iterations (usually $m\gg N$ and $n\gg N$)
\end{center}
\end{table*}

\subsection{Distributed Multisplitting method}
\label{subsec:Multisplitting}
The first instances of iterative methods for solving linear systems involve the well known four main stationary iterative methods, Jacobi method, Gauss-Seidel method, successive overrelaxation method (SOR), and symmetric successive overrelaxation method (SSOR)~\cite{HY-AIM1981}. To parallelize the stationary iterative methods, space decomposition methods are employed to partition the matrix $A$ into blocks (block Jacobi or block SOR) as well as the original problem into smaller local problems~\cite{FR-JCAM1999}. Renaut~\cite{R-NLAA1998} also proposed a parallel multisplitting (MS) solution of the least-squares problem where the solutions to the local problems are recombined using weighting matrices.

Stationary iterative methods are based on a single splitting, $A=L-U$ which is well known~\cite{OR-BOOK2000}. Multisplittings (MS) generalize the splitting to take advantage of the computational capabilities of parallel computers. A multisplitting of A is defined as follows,\\

\noindent{\bf Definition.} Linear multisplitting (LMS). Given a matrix $A\in\mathbb{R}^{n\times n}$ and a collection of matrices $L^{(u)},U^{(u)},E^{(u)}\in\mathbb{R}^{n\times n}$,$u=1:N$, satisfying
\begin{itemize}
 \item $A=L^{(u)}-U^{(u)}$,
 \item $L^{(u)}$ is regular,
 \item $E^{(u)}$ is a non-negative diagonal matrix, and $\sum^N_{u=1}E^{(u)}=I$.
\end{itemize}
Then the collection of triples $(L^{(u)},U^{(u)},E^{(u)})$ is called a multisplitting of $A$ and the LMS method is defined by the iteration:
$$x^{k+1}=\sum^N_{u=1}E^{(u)}(L^{(u)})^{-1}(U^{(u)}x^k+b), k=1,\ldots$$
The advantage of this method is that at each iteration there are $N$ independent problems of the kind
$$L^{(u)}y^k_u=U^{(u)}x^k+b,u=1:N$$
where $y^k_u$ represents the solution to the $u$th local problem. Then we can assign each local problem to one node in the network and the communication is only required to produce the update.

\begin{figure}[htb!]
\centering
\includegraphics[width=3.4in]{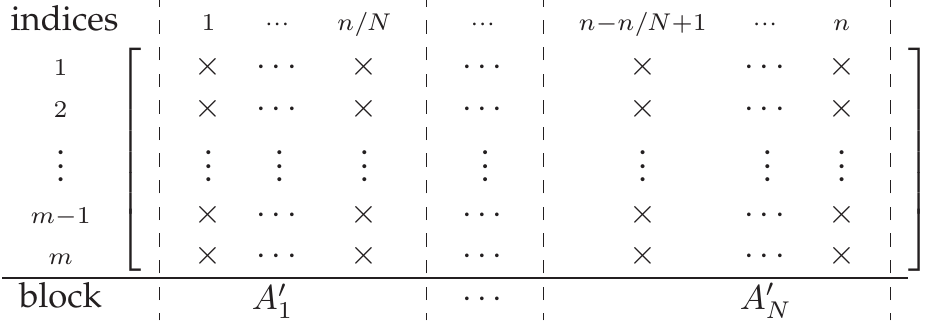}
\caption{Column partition of matrix $A$}
\label{fig:colpart}
\end{figure}

Notice that it is different from our previous assumption, in multisplitting the matrix $A$ is partitioned into blocks of columns consistently with the decomposition of $x$ into blocks as $A=(A'_1,A'_2,\ldots,A'_N)$, where each $A'_u\in\mathbb{R}^{m\times n_u}$ and $n_u=n/N$ as shown in Figure~\ref{fig:colpart} (as we discussed in row partition, the partition does not have to be equally, $n_u=n/N$ is for simplicity of analysis). For example, node $v_1$ holds $A'_1$ and part of vector $x$, $\{x_1,\ldots,x_{n/N}\}$. To avoid ambiguity, in this paper let $A'_u$ denote the column block on node $v_u$ and $A^T_u$ denote the transpose of the row block $A_u$ on node $v_u$.

With the column decomposition $Ax=\sum^N_{u=1}A'_ux_u$ can be replaced by the subproblems
$$\min_{y\in\mathbb{R}^{n_u}}||A'_uy-b_u(x)||_2$$
where $b_u(x)=b-\sum_{u'\neq u}A'_{u'}x_{u'}=b-Ax+A'_ux_u$. Each of the local problems is also a linear least-squares problem, which can be solved by QR decomposition of the matrix $A'_u$. Let $x^k=(x^k_1,x^k_2,\ldots,x^k_N)$ be the solution at iteration $k$, then the solution at iteration $k+1$ can be constructed from the local problems
$$\min_{y^{k+1}_u\in\mathbb{R}^{n_u}}||A'_uy^{k+1}_u-b_u(x^k)||_2, 1\leq u\leq N$$
according to
$$x^{k+1}=\sum^N_{u=1}\alpha^{k+1}_u\bar{x}^{k+1}_u$$
The updated local solution to the global problem is given by
$$\bar{x}^{k+1}_u=(x^k_1,x^k_2,\ldots,x^k_{u-1},y^{k+1}_u,x^k_{u+1},\ldots,x^k_N)$$
where the non-negative weights satisfy $\sum^N_{i=1}\alpha^{k+1}_u=1$. For the $u$th column block of matrix $A$,
\begin{eqnarray*}
x^{k+1}_u&=&x^{k}_u + \alpha^{k+1}_u(y^{k+1}_u-x^{k}_u)\\
&=&x^{k}_u + \alpha^{k+1}_u\delta^{k+1}_u
\end{eqnarray*}
where now $\alpha^{k+1}_u\delta^{k+1}_u$ is the step taken on partition $u$. The update of $b_u(x^k)$ can be expressed as
$$b_u(x^{k+1})=b_u(x^k)-\sum^N_{\substack{u'=1\\u'\neq u}}\alpha^{k+1}_{u'}B^{k+1}_{u'}$$
here we define $B^{k+1}_{u'}=A'_{u'}\delta^{k+1}_{u'}$.

\begin{algorithm}[htb!]
\caption{D-MS method}
\label{alg:lsms}
\leftline{Each node $v_u$ follows the same routines below}
\begin{algorithmic}[1]
\STATE calculate $Q_uR_u=A'_u$
\STATE $y^0_u=x^0_u,k=0,\alpha^0_u=0,\alpha^k_u=\alpha=\frac{1}{N},b_u(x^0)=b$.
\WHILE{not converged}
\STATE $k=k+1$
\STATE $B_u=A'_u\delta_u$
\STATE \textcolor[rgb]{0.50,0.50,0.50}{\bf flood $B_u$ to all other nodes in network}
\STATE update $b_u$
\STATE solve $\min_{y\in\mathbb{R}^{n_u}}||A'_uy-b_u(x)||_2$ and get $y_u$
\STATE $\delta_u=y_u-x_u$
\STATE $x_u=x_u+\alpha\delta_u$
\ENDWHILE
\end{algorithmic}
\end{algorithm}

\begin{figure*}[htb!]
\centering
\includegraphics[width=4.8in]{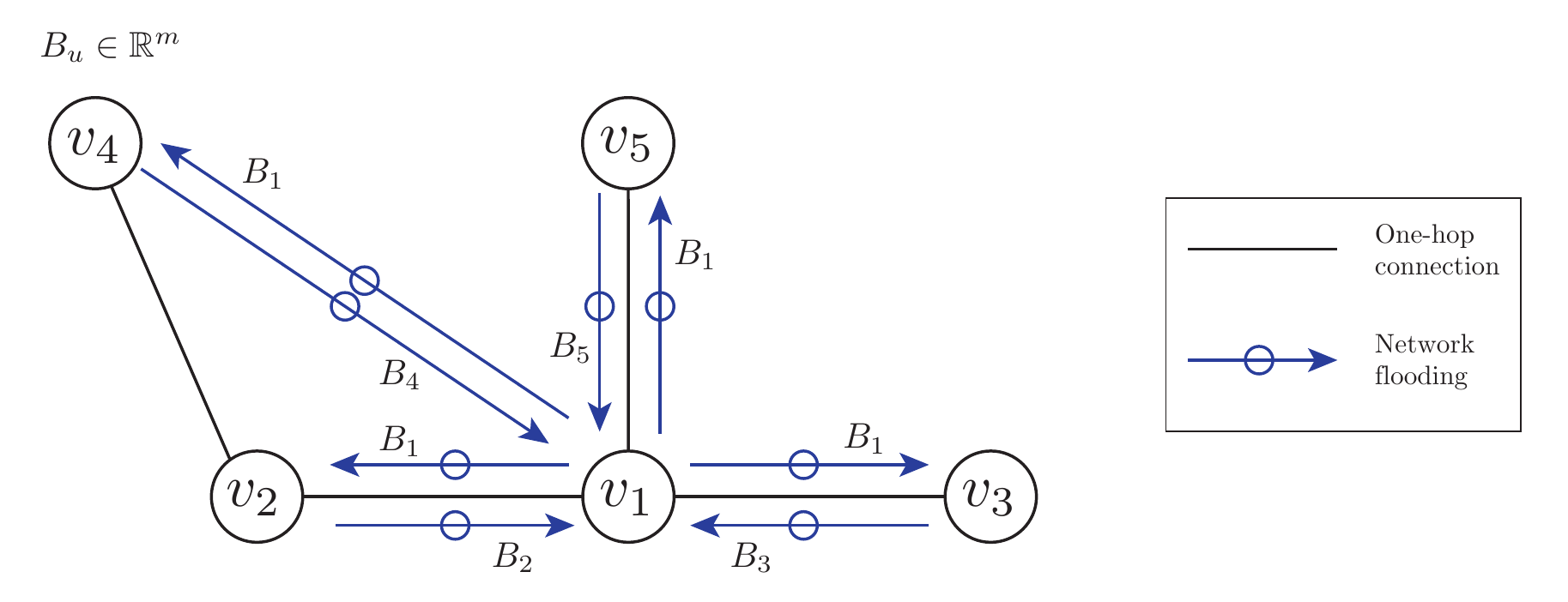}
\caption{Communication pattern of step 7 in algorithm~\ref{alg:lsms} D-MS method}
\label{fig:comm2}
\end{figure*}

Algorithm~\ref{alg:lsms} gives the description of the Distributed Multisplitting (D-MS) method. Notice that the communication only happens in line 6, and this is a network flooding. To illustrate the communication patter we use a mesh network example as shown in Figure~\ref{fig:comm2} to show the communications in D-MS method. In the example we show the messages sent from and received by node $v_1$ in the algorithm, similar examples will be illustrated in other methods of this paper.

At the beginning of the algorithm, each node holds a column partition of matrix $A$, since in section~\ref{sec:netmod} we assume that the sensed information of each node usually forms a row partition of matrix $A$. To directly apply this algorithm on the system with non-symmetric matrix $A$, a one time communication is required in the network to exchange the information of matrix $A$ and let node $v_u$ has $A'_u$. Since many applications results a symmetric matrix $A$ where $A_u=A'_u$, this communication is not necessary all the time. \\


\noindent{\bf Communication cost.} In line 6 of Algorithm~\ref{alg:lsms}, for each iteration node $v_u$ needs to flood the vector $B_u$ to all other nodes in the network and the length of vector $B_u$ is $m$. Although the matrix $A$ may be sparse, $B_u$ is a dense vector, thus we use the length of the vector to denote the communication cost without loss of generality. For example, in Figure~\ref{fig:comm2}, node $v_1$ flood $B_1$ of dimension $m$ (rows of matrix $A$), and all other nodes in the network will receive $B_1$ to update the local $b_u$ as well as the local solutions. Since the network size is $N$, the communication cost for each flood from one node is $mN$. So the communication cost for all the nodes in the network to flood once is $mN^2$. Suppose that after $k$ iterations the algorithm converges, then the total communication cost is $kmN^2$.

\noindent{\bf Time-to-completion.} The only communication pattern in Algorithm~\ref{alg:lsms} is network flooding, so for the time-to-completion, each node $v_u$ needs to flood $B_u$ to the network and receive the $B_{u'}$ from all other nodes $u'\in N$ and $u'\neq u$. From the assumption in section~\ref{sec:netmod}, the communication delay of one unit data delivery between direct neighbors would be one unit time and each node only has one radio. First, the maximum delay of transmitting $B_u$ from $u$ to a node $u'$ in the network is $\log N$. But from the receiver side, since the node only has one radio, $v_u$ needs at least $N-1$ unit time to receive all $B_{u'}$s from other nodes. Since $N-1\geq\log N$, considering the length of $B_u$ the time-to-completion of one iteration is $m(N-1)$ and the total time-to-completion is then $km(N-1)$.


\subsection{Distributed Modified Conjugate Gradient Least-Squares Method}
\label{subsec:MCGLS}
One common and often efficient approach to solve least-squares problem is minimizing by solving the normal equations, because $A^TA$ is symmetric and positive definite and it can be solved by using conjugate gradient method which was originally developed by Lanczos~\cite{L-JRNBS1952} and Hestenes and Stiefel~\cite{HS-JRNBS1952}. The resulting method, CGLS, is often used as the basic iterative method to solve the least-squares problems~\cite{BES-JMAA1998}. On parallel architectures, the basic computation operations of iterative methods are usually: inner products, vector updates, matrix vector products. Yang and Brent~\cite{YB-TJS2004} describe a modified conjugate gradient least-squares (MCGLS) method to reduce inner products global synchronization points and improve the parallel performance. This can also be potentially distributed over the mesh network. This algorithm is based on the MCGLS method to reduce inner products global synchronization points, then improve the parallel performance accordingly. In the MCGLS method, there are two ways of improvement. One is to assemble the results of a number of inner products collectively and another is to create situations where communication can be overlapped with computation.

\begin{algorithm*}[htb!]
\caption{The MCGLS and D-MCGLS method}
\label{alg:cgls}
\noindent\begin{minipage}{\textwidth}
  \begin{minipage}{.38\textwidth}
    \label{alg:algcg1}
    \begin{algorithmic}[1]
    \STATE Let $x^0$ be an initial guess
    \STATE $r^0=b-Ax^0$
    \STATE {\color{black}$s^0=p^0=A^Tr^0$}
    \STATE {\color{black}$\gamma^0=(s^0,s^0)$}
    \STATE $k=0$
    \WHILE{not converged}
    \STATE $k=k+1$
    \STATE $p^k=s^k+\beta^{k-1}p^{k-1}$
    \STATE {\color{black}$q^k=Ap^k$}
    \STATE {\color{black}$\delta^k=(q^k,q^k)$}
    \STATE $x^k=x^{k-1}+\alpha^{k-1}p^{k-1}$
    \STATE $\alpha^k=\gamma^k/\delta^k$
    \STATE $r^{k+1}=r^k-\alpha^kq^k$
    \STATE {\color{black}$s^{k+1}=A^Tr^{k+1}$}
    \STATE {\color{black}$\gamma^k=(s^{k+1},s^{k+1})$}
    \STATE $x^{k+1}=x^k+\alpha^kp^k$
    \STATE $\beta^k=\gamma^{k+1}/\gamma^k$
    \ENDWHILE
    \end{algorithmic}
  \end{minipage}
  \begin{minipage}{.60\textwidth}
    \leftline{Each node $v_u$ follows the same routines below}
    \label{alg:algcg2}
    \begin{algorithmic}[1]
    \STATE initial $x^0$ (same for all nodes)
    \STATE $r^0_u=b_u-A_{u}x^0$ ($\in\mathbb{R}^{m_u}$), \textcolor[rgb]{0.50,0.50,0.50}{\bf flood $r^0_u$}
    \STATE $r^0=\sum^N_{u=1}r^0_u$, $s^0_u=p^0_u=A'_{u}r^0$ ($\in\mathbb{R}^{n_u}$)
    \STATE $\gamma^0_u=(s^0_u,s^0_u)$, \textcolor[rgb]{0.50,0.50,0.50}{\bf flood $\gamma^0_u$}, $\gamma^0=\sum^N_{u=1}\gamma^0_u$
    \STATE $k=0$
    \WHILE{not converged}
    \STATE $k=k+1$
    \STATE $p^k_u=s^k_u+\beta^{k-1}p^{k-1}_u$ ($\in\mathbb{R}^{n_u}$), \textcolor[rgb]{0.50,0.50,0.50}{\bf flood $p^k_u$}
    \STATE $p^k=\sum^N_{u=1}p^k_u$, $q^k_u=A_{u}p^k$ ($\in\mathbb{R}^{m_u}$)
    \STATE $\delta^k_u=(q^k_u,q^k_u)$, \textcolor[rgb]{0.50,0.50,0.50}{\bf flood $\delta^k_u$}, $\delta^k=\sum^N_{u=1}\delta^k_u$
    \STATE $x^k_u=x^{k-1}_u+\alpha^{k-1}p^{k-1}_u$ ($\in\mathbb{R}^{n_u}$)
    \STATE $\alpha^k=\gamma^k/\delta^k$
    \STATE $r^{k+1}_u=r^k_u-\alpha^kq^k_u$ ($\in\mathbb{R}^{m_u}$), \textcolor[rgb]{0.50,0.50,0.50}{\bf flood $r^{k+1}_u$}
    \STATE $r^{k+1}=\sum^N_{u=1}r^{k+1}_u$, $s^{k+1}_u=A'_{u}r^{k+1}$ ($\in\mathbb{R}^{n_u}$)
    \STATE $\gamma^k_u=(s^{k+1}_u,s^{k+1}_u)$, \textcolor[rgb]{0.50,0.50,0.50}{\bf flood $\gamma^k_u$}, $\gamma^k=\sum^N_{u=1}\gamma^k_u$
    \STATE $x^{k+1}_u=x^k_u+\alpha^kp^k_u$ ($\in\mathbb{R}^{n_u}$)
    \STATE $\beta^k=\gamma^{k+1}/\gamma^k$
    \ENDWHILE
    \end{algorithmic}
  \end{minipage}
\end{minipage}
\end{algorithm*}

This section gives the description and analysis of the distributed modified conjugate gradient least-squares (D-MCGLS) method. When the matrix $A$ has full rank, there is a unique solution $\hat{x}$ for the system of normal equations, let $\hat{r}=b-A\hat{x}$ be the corresponding residual. Give a initial vector $x^0$ the conjugate gradient algorithm generates approximations $x^k$ in the subspace
$$x^k\in x^0+\mathcal{K}_k(A^TA,s_0), s_0=A^T(b-Ax^0)$$
where $\mathcal{K}_k(A^TA,s^0)$ is the Krylov subspace
$$\mbox{span}\{A^Ts^0,(A^TA)A^Ts^0,\ldots,(A^TA)^{(k-1)}A^Ts^0\}$$
The iterates are optimal in the sense that for each $k$, $x^k$ minimizes the error functional
$$E_{\mu}(x^k)=(\hat{x}-x^k)^T(A^TA)^{\mu}(\hat{x}-x^k)$$
Only the values $\mu=0,1,2$ are of practical interest. By using $A(\hat{x}-x^k)=b-\hat{r}-Ax^k=r^k-\hat{r}$,
\begin{equation*}
 E_{\mu}(x^k)=\left\{
  \begin{array}{ll}
   ||\hat{x}-x^k||^2, & \mu=0;\\
   ||\hat{r}-r^k||^2=||r^k||^2-||\hat{r}||^2, & \mu=1;
  \end{array}
 \right.
\end{equation*}
We consider here only the case $\mu=1$, namely CGLS, which is of most practical interest with the best numerical accuracy. The MCGLS algorithm originally given in~\cite{YB-TJS2004} is shown on the left side in algorithm~\ref{alg:cgls}, the corresponding distributed version D-MCGLS is shown on the right side in algorithm~\ref{alg:cgls}.

\begin{figure*}[htb!]
\centering
\includegraphics[width=4.8in]{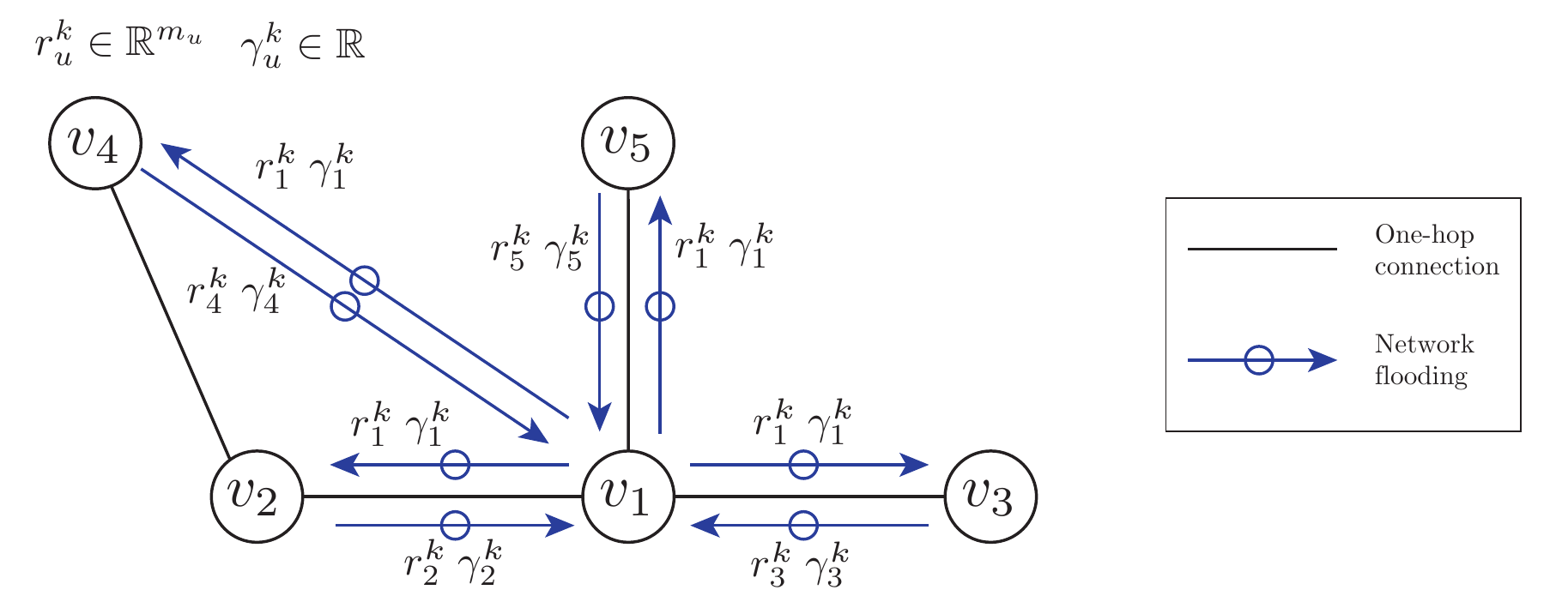}
\caption{Communication Pattern of step 2, 4, 13 and 15 in Algorithm~\ref{alg:cgls} D-MCGLS method }
\label{fig:comm3-1}
\end{figure*}

\begin{figure*}[htb!]
\centering
\includegraphics[width=4.8in]{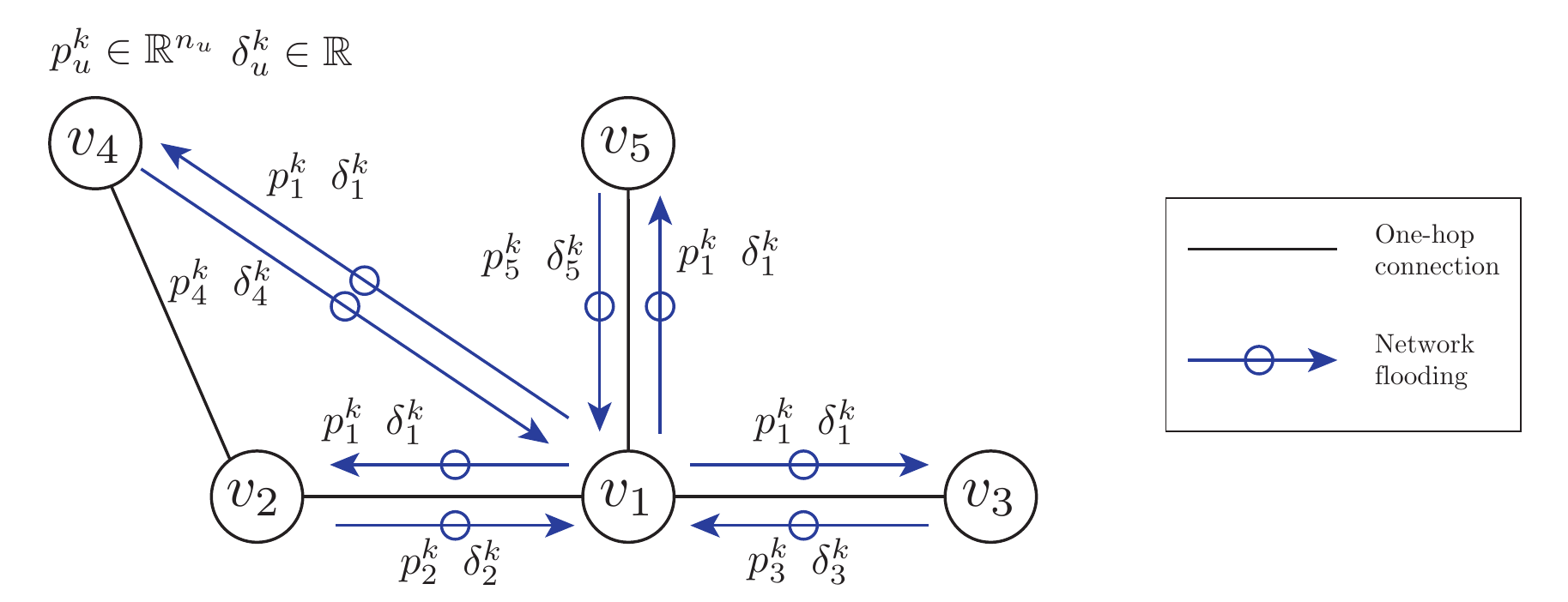}
\caption{Communication Pattern of step 8 and 10 in Algorithm~\ref{alg:cgls} D-MCGLS method}
\label{fig:comm3-2}
\end{figure*}

\noindent{\bf Communication cost.} To distribute the MCGLS algorithm, each node in the network has the row partition $A_u$ of matrix $A$ as well as the column partition $A^T_u$ of $A$ (as shown in section~\ref{subsec:DLMS} where one time communication on columns information of $A$ may happen if $A$ is not symmetric and the columns information exchange cost analysis is omitted here). Then setp 2, 4, 8, 10, 13 and 15 require communications (2, 4 are just the initial steps of 13, 15 which are one time communications).

From our assumption, at the beginning of the algorithm, each node $v_u$ has $m_u$ consecutive rows and $n_u$ consecutive columns of $A$, i.e., each node has part of matrix $A$ and part of matrix $A^T$. Then for each node $v_u$, it handles part of vector $s^k_u, p^k_u, q^k_u$ and $x^k_u$. Note that the dimensions of $s^k_u, p^k_u$ and $q^k_u, x^k_u$ are different, $s^k_u, p^k_u$ is of dimension $m_u\times n$ ($\sum^N_{u=1}m_u=m$) while $q^k_u, x^k_u$ is of dimension $m\times n_u$ ($\sum^N_{u=1}n_u=n$).

Next we give an analysis on the communication cost of the D-MCGLS method. First node $v_u$ compute a partial residual vector $r^0_u$ of dimension $m_u$. Then node $v_u$ needs to flood $r^0_u$ and each node needs to compute $r_0$ in line 2 by sum up the $r^0_u$ from all other nodes as shown in Figure~\ref{fig:comm3-1}. For example, node $v_1$ receives different $r^0_u$s and $\gamma^0_u$s from all other nodes, then it can calculate $r_0=\sum^N_{u=1}r^0_u$ as well as $\gamma_0=\sum^N_{u=1}\gamma^0_u$. Here the communication cost for each flood from node $v_u$ is of $Nm_u$, the total communication cost in the network is then $N\sum^{N}_{u=1}m_u=mN$. In line 4, we need to compute the inner product of vector $s^0$. Since each node $v_u$ has a partial vector $s^0_u$, node $v_u$ can compute the partial product value $\gamma^0_u$ and flood the value to all other nodes. Then each node will get $\gamma^0$ by sum up all the partial values. The communication cost for flood $\gamma^0_u$ in the network is $N^2$ since $\gamma^0_u$ is a scalar value.

After initialization, the computation starts on each node, after computing $p^k_u$ in line 8, each node needs to flood this to the network so that all the nodes can compute $p^k$ to update $q^k_u$. The communication cost for each flood on node $v_u$ is of $Nn_u$, the total communication cost in the network is then $N\sum^{N}_{u=1}n_u=Nn$. To compute the inner product of $q^k_u$, the communication pattern and cost is the same as the above analysis for $\gamma^0$, $N^2$ as shown in Figure~\ref{fig:comm3-2}. Then the communication pattern in line 13 and 15 are exactly the same as in step 2 and 4. Suppose that after $k$ iterations, the algorithm converges, the total communication cost of the network is $(k+1)(m+N)N+k(n+N)N$.

\noindent{\bf Time-to-completion.} Following the analysis in previous section. The network flooding communication in Figure \ref{fig:comm3-1} results the time-to-completion $k(m+1)(N-1)$, and the flooding communication in Figure \ref{fig:comm3-2} results the time-to-completion $k(n+1)(N-1)$ in the network. So the total time-to-completion of this algorithm is $k(m+n+2)(N-1)$.

\subsection{Distributed Least Mean Squares Method}
\label{subsec:DLMS}


Schizas, Mateos and Giannakis~\cite{MSG-EURASIPJASP2010, MSG-AAC2007, SMG-ACSSC2007, SRG-TSP20081, SGRR-TSP20084, SMG-TSP2009, MG-TOSP2012} introduce the Distributed Least Mean Square (D-LMS) algorithms. This algorithm let each node maintain its own local estimation and, to reach the consensus, exchange the local estimation only within its neighbors. The advantage of the methods like D-LMS and D-CE in signal processing is that only local information exchange is required. The problem is that these methods may converge slow~\cite{TS-TSP2012} in a large-scale network.

In their discussion, the wireless sensor network is deployed to estimate a signal vector $x^*\in \mathbb{R}^{n\times 1}$, where $k=0,1,2,\ldots$ denote the time instants, each node $v_u$ has a regression vector $A_u(k)\in \mathbb{R}^{n\times 1}$ and there is a observation $b_u(k)$ on time $k$, both of them are assumed to have zero mean. One global vector $b(k):=[b_1(k)\ldots b_N(k)]^T\in \mathbb{R}^{N\times1}$ is used for all the observations on $N$ nodes in the network. $A(k):=[A_1(k)\ldots A_N(k)]^T\in\mathbb{R}^{N\times n}$ is the regression vectors combined over the network, the global LMS estimator is then described as
\begin{eqnarray*}
\hat{x}(k)&=&\arg\min_xE[||b(k)-A(k)x||^2]\\
&=&\arg\min_x\sum^N_{u=1}E[(b_u(k)-A^T_u(k)x)^2]
\end{eqnarray*}

Let $\{x_u\}^N_{u=1}\in\mathbb{R}^n$ represent the local estimation of the global variable $x$ one node $v_u$ (each node has its own estimation of the signal vector). In conjunction with these local variables, consider the convex constrained minimization problem
\begin{eqnarray*}
\{\hat{x}_u(k)\}^N_{u=1}&=&\arg\min_x\sum^N_{u=1}E[(b_u(k)-A^T_u(k)x_u)^2]\\
&&\mbox{s.t. } x_u = x_{u'}, u\in N, u'\in\mathcal{N}_u
\end{eqnarray*}
where $\mathcal{N}_u$ is the neighbor set of node $v_u$.

The equality constraints above only involve the local estimations of the neighbors of each node and forces an agreement among each node's neighbors. Since we assumed that the network is connected, the constraints above will introduce a consensus in the network. Then we can finally have $x_u=x_{u'}$ for all $u,u'\in N$. So we find that the distributed estimation problem is equivalent to the original problem in the sense that their optimal solutions coincide such as $\hat{x}_u(k)=\hat{x}(k)$, for all $u\in N$.

To construct the distributed algorithm, the authors resort to the AD-MoM algorithm, and get the following two equations for estimation updating,
\begin{eqnarray*}
v^{u'}_u(k)&=&v^{u'}_u(k-1)+\frac{c}{2}(x_u(k)-(x_{u'}(k)+\eta^{u'}_u(k))),\\
&&u'\in\mathcal{N}_u\\
x_u(k+1)&=&x_u(k)+\mu_u\big[2A_u(k+1)e_u(k+1)-\\
&&\sum_{u'\in\mathcal{N}_u}(v^{u'}_u(k)-(v^{u}_{u'}(k)+\bar{\eta}^{u'}_u(k)))-\\
&&c\sum_{u'\in\mathcal{N}_u}(x_u(k)-(x_{u'}(k)+\eta^{u'}_u(k)))\big]
\end{eqnarray*}
where $\mu_u$ is a constant step-size and $e_u(k+1):=b_u(k+1)-A^T_u(k+1)x_u(k)$ is the local a priori error. ${\bf\eta}^{u'}_u(k)$ and ${\bf\bar{\eta}}^{u'}_u(k)$ denote the additive communication noise present in the reception of $x_{u'}(k)$ and $v^{u}_{u'}(k)$. Algorithm~\ref{alg:dlms} gives the description of the distributed least mean square algorithm. In detail, during time instant $k+1$ node $v_u$ receives the local estimates $\{x_{u'}(k)+\eta^{u'}_u(k)\}_{u'\in\mathcal{N}_u}$ and plugs them into the equations above to evaluate $v^{u'}_u(k)$ for $u'\in\mathcal{N}_u$. Each one of the updated local Lagrange multipliers $\{v^{u'}_u(k)\}_{u'\in\mathcal{N}_u}$ is subsequently transmitted to the corresponding neighbor $u'\in\mathcal{N}_u$. then upon reception of $\{v^{u}_{u'}(k)+\bar{\eta}^{u'}_u(k))\}_{u'\in\mathcal{N}_u}$, the multipliers are jointly used along with $\{x_{u'}(k)+\eta^{u'}_u(k))\}_{u'\in\mathcal{N}_u}$ and the newly acquired local data $\{b_u(k+1),A_u(k+1)\}$ to obtain $x_u(k+1)$ via the above equations. The $(k+1)$-st iteration is concluded after node $v_u$ broadcast $x_u(k+1)$ to its neighbors.

\begin{algorithm}[htb!]
\caption{D-LMS Method}
\label{alg:dlms}
\leftline{Each node $v_u$ follows the same routines below}
\begin{algorithmic}[1]
\STATE Arbitrarily initialize $\{x_u(0)\}^N_{i=1}$ and $\{v^{u'}_u(-1)\}^{u'\in\mathcal{N}_u}_{i\in N}$
\WHILE{not converged}
\STATE \textcolor[rgb]{0.50,0.50,0.50}{\bf Broadcast $x_u(k)$ to neighbors in $\mathcal{N}_u$}
\STATE Update $\{v^{u'}_u(k)\}_{u'\in\mathcal{N}_u}$
\STATE \textcolor[rgb]{0.50,0.50,0.50}{\bf Transmit $v^{u'}_u(k)$ to each $u'\in\mathcal{N}_u$}
\STATE Update $x_u(k+1)$
\ENDWHILE
\end{algorithmic}
\end{algorithm}

\begin{figure*}[htb!]
\centering
\includegraphics[width=4.8in]{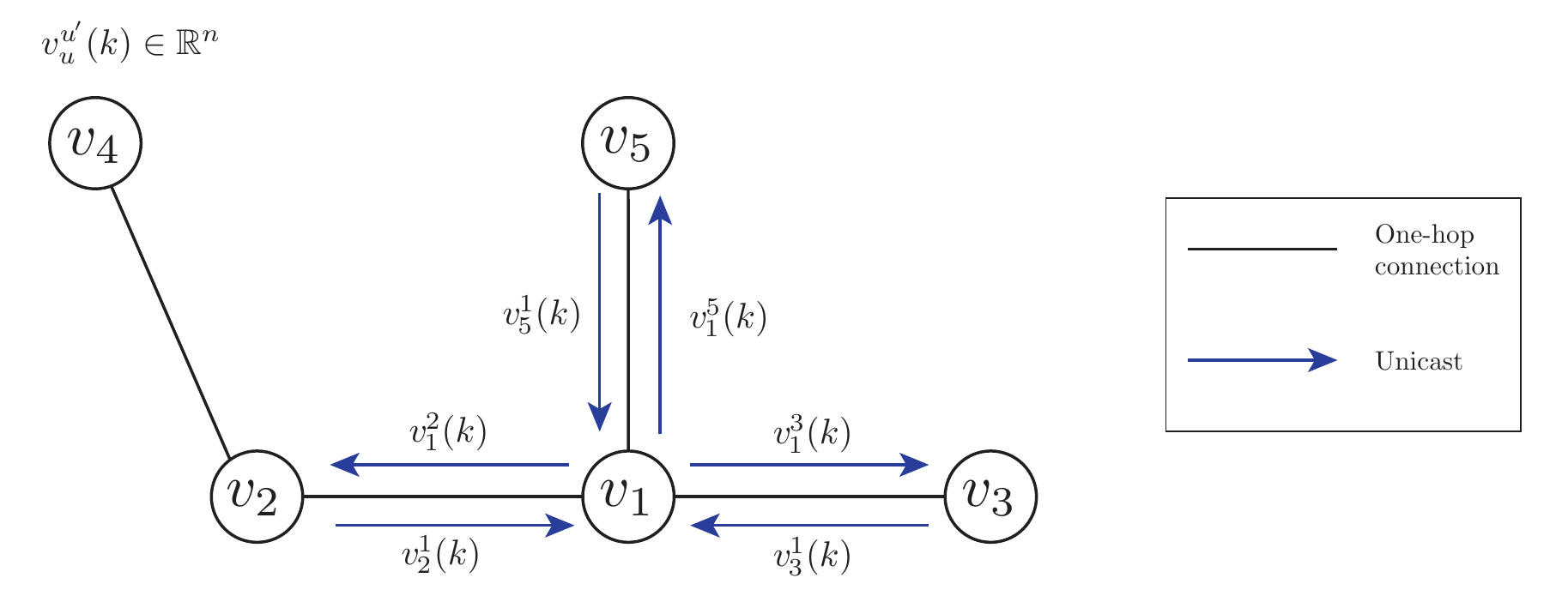}
\caption{Communication Pattern of step 5 in Algorithm~\ref{alg:dlms} D-LMS method}
\label{fig:comm4}
\end{figure*}

\noindent{\bf Communication cost.} The algorithm framework is simple and only step 3 and 5 involves communication. Applied to system $Ax=b$, vectors $x_u(k)$ and $v^{u'}_u(k)$ are both of length $n$ (columns of $A$). As shown in Figure~\ref{fig:comm4}, e.g., in step 3 node $v_1$ needs to transmit $x_u(k)$ to all its neighbors. In step 5 $v_1$ needs to transmit different $v^{u'}_u(k)$ to different neighbors. Suppose that the average degree of network is $D_{avg}$, in each iteration the communication cost of one node is of $n(D_{avg}+1)$, so the communication of the network is $nN(D_{avg}+1)$. Suppose that after $k$ iterations (the iteration number might be greater than the time instants, so after the $k$-th sample on node $v_u$ is involved, the first sample is used as the $(k+1)$-st sample), the algorithm converges, then the total communication cost is $knN(D_{avg}+1)$.

\noindent{\bf Time-to-completion.} In step 3 of the algorithm, node $u$ needs to broadcast $x_u(k)$ to all its neighbors. From the receiver side, each node need to receive different update from all its neighbors, then the delay of the whole network depends on the maximum node degree of the network since the algorithm is synchronous, this delay is $nD_{max}$. In step 5, node $u$ needs to send different $v^{u'}_u(k)$ to different neighbors, the delay is also $nD_{max}$. The total communication delay is then $2knD_{max}$.

\subsection{Distributed Recursive Least-Squares Method}
\label{subsec:DRLS}
Sayed and Lopes~\cite{SL-ACSSC2006} developed a distributed least-squares estimation strategy by appealing to collaboration techniques that exploit the space-time structure of the data, achieving an exact recursive solution that is fully distributed. This Distributed Recursive Least-Squares (D-RLS) strategy is developed by appealing to collaboration techniques to achieve an exact recursive solution. It requires a cyclic path in the network to perform the computation node by node. The advantage of this method is the iteration number is fixed (the network size) for a give set of data to solve a least-squares problem, but the problem is a large dense matrix needs to be exchanged between nodes.

The details and analysis of D-RLS strategy are given in this section, Algorithm~\ref{alg:rls} gives the classic RLS procedure~\cite{M-DSPF2009}
\begin{algorithm}[htb!]
\caption{Recursive Least-Squares Procedure}
\label{alg:rls}
Initial: $x^{-1}=\bar{x}$ and $P^{-1}=I$
\begin{algorithmic}[1]
\FOR{$k\geq0$}
\STATE $x^k=x^{k-1}+g^k[b(k)-A^T(k)x^{k-1}]$
\STATE $g^k=\frac{\lambda^{-1}P^{k-1}A(k)}{1+\lambda^{-1}A^T(k)P^{k-1}A(k)}$
\STATE $P^k=\lambda^{-1}[P^{k-1}-g^kA^T(k)P^{k-1}]$
\ENDFOR
\end{algorithmic}
\end{algorithm}
where $P^k\in\mathbb{R}^{n\times n}$.

To distribute the exact algorithm for estimating the vector $x$ in the network of $N$ nodes, each node $v_u$ has access to regressors and measurement data $A_u(k)$ and $b_u(k)$, $u=1,\ldots,N$, where $b_u(k)\in\mathbb{R}$ and $A_u(k)\in\mathbb{R}^n$. At each time instant $k$, the network has access to space-time data
$$b(k)=\left[
\begin{array}{c}
 b_1(k)\\
 b_2(k)\\
 \vdots\\
 b_N(k)
\end{array}
\right]
\mbox{ and }
A(k)=\left[
\begin{array}{c}
 A_1(k)\\
 A_2(k)\\
 \vdots\\
 A_N(k)
\end{array}
\right]$$
Here $b(k)$ and $A(k)$ are snapshot matrices revealing the network data status at time $k$. We collect all the data available up to time $k$ into global matrices $b$ and $A$
$$b=\left[
\begin{array}{c}
 b(0)\\
 \hline
 b(1)\\
 \hline
 \vdots\\
 \hline
 b(k)
\end{array}
\right]
\mbox{ and }
A=\left[
\begin{array}{c}
 A(1)\\
 \hline
 A(2)\\
 \hline
 \vdots\\
 \hline
 A(k)
\end{array}
\right]$$
Notice that here is equivalent to solve the least-squares problem of $Ax=b$ by partition $A$ and $b$ row-wise and each node has one partition of consecutive rows of $A$ and $b$. Applying RLS algorithm here is different to the D-LMS estimator since it gives the least-squares solution of the whole block of data $Ax=b$. So in the distributed RLS algorithm to solve a normal least-squares problem, we use $A_u(k)$ indicate the row block on node $v_u$ but not only one vector collected at time instant $k$ (we can treat it as all the data collected till time $k$).

By assuming an incremental path is defined across the network cycling from node $v_1$ to $v_2$ and so forth, until node $v_N$. The RLS algorithm can be rewritten as a distributed version in algorithm~\ref{alg:drls}~\cite{SL-ACSSC2006}.\\
\begin{algorithm}[htb!]
\caption{D-RLS Method}
\label{alg:drls}
$\psi^{(k)}_0=x^{k-1}$,$P_{0,k}=\lambda^{-1}P^{k-1}$
\begin{algorithmic}[1]
\FOR{$u=1:N$, node $v_u$}
\STATE $e_u(k)=b_u(k)-A_u(k)\psi^{(k)}_{u-1}$
\STATE $\psi^{(k)}_u=\psi^{(k)}_{u-1}+\frac{P_{u-1,k}}{\gamma^{-1}_u+A^T_u(k)P_{u-1,k}A_u(k)}A_u(k)e_u(k)$
\STATE $P_{u,k}=P_{u-1,k}-\frac{P_{u-1,k}A_u(k)A^T_u(k)P_{u-1,k}}{\gamma^{-1}_u+A^T_u(k)P_{u-1,k}A_u(k)}$
\IF{$u\neq N$}
\STATE \textcolor[rgb]{0.50,0.50,0.50}{\bf $v_u$ send $\{\psi^{(k)}_u,P_{u,k}\}$ to node $v_{u+1}$}
\ENDIF
\ENDFOR
\end{algorithmic}
\end{algorithm}

\begin{figure*}[htb!]
\centering
\includegraphics[width=4.8in]{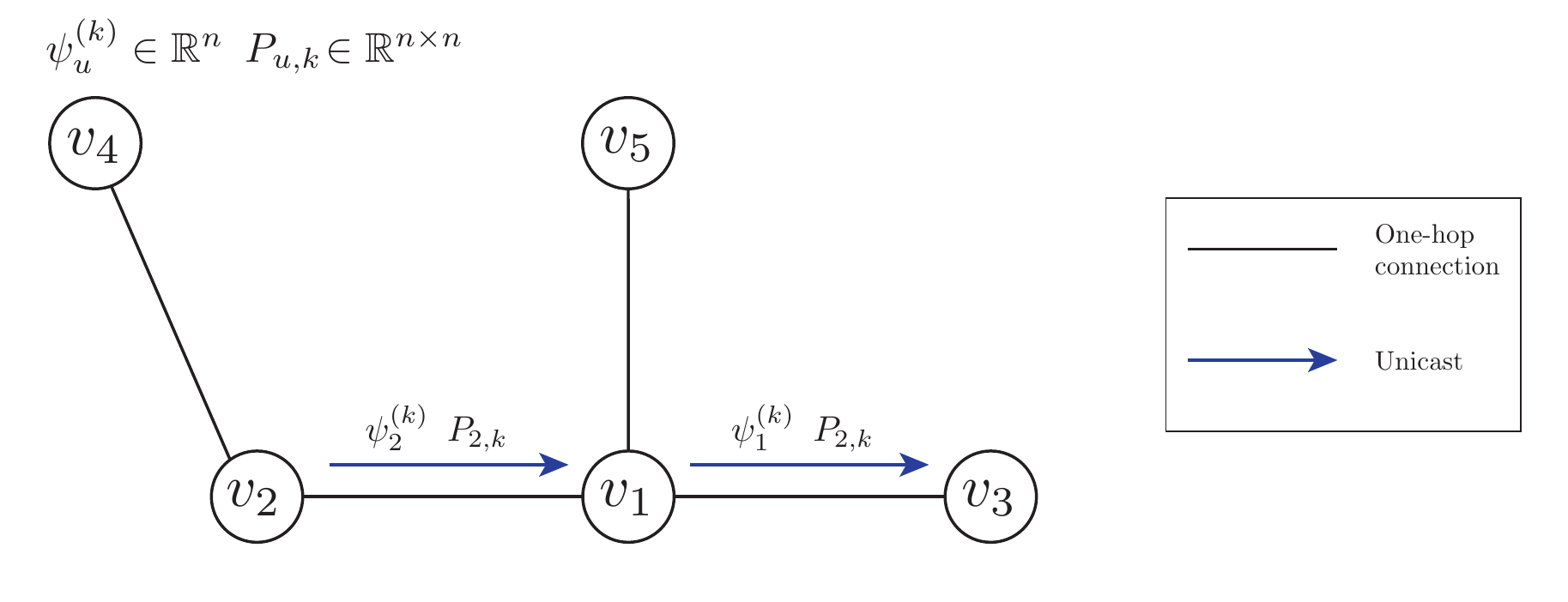}
\caption{Communication Pattern of D-RLS method}
\label{fig:comm7}
\end{figure*}

\noindent{\bf Communication cost.} In the distributed RLS algorithm, the communication is in step 6. Each node shares with its successor node in the cycle path of network the quantities $\{\psi^{(k)}_u,P_{u,k}\}$ where $\psi^{(k)}_u\in\mathbb{R}^n$ and $P_{u,k}\in\mathbb{R}^{n\times n}$. For example, in Figure~\ref{fig:comm7} node $v_1$ receives the message from $v_2$ and sends it to $v_3$. So in each iteration, the communication cost only happens in one node and it is $n+n^2$. Since the algorithm can converge after one cycle in the network, then the total communication cost is $(n+n^2)(N-1)$. Note that a Hamiltonian path is required by this algorithm, to find such a path, extra communication is required. This is another problem and out of the scope of the analysis in this paper, we omit this cost here.

\noindent{\bf Time-to-completion.} In distributed RLS algorithm, it is easy to see that the delay in one step is $n+n^2$, there are totally $N-1$ steps in the algorithm, so the total time-to-completion is $(n+n^2)(N-1)$.

\section{Conclusion}
\label{sec:conc}
In this paper, we surveyed some of the developments in distributed iterative methods and parallel iterative methods which can be potentially applied to solve least-squares problems in the mesh network. We have covered the traditional iterative methods for solving linear systems including the relaxation methods, the conjugate gradient methods and the row action methods. One algorithm from each category is selected to be described in details that how to apply them to solve least-squares problem in mesh network. Besides, we also surveyed some of the consensus and diffusion based strategies for parameter estimation in signal processing in the network. Compared to the traditional iterative methods, the consensus and diffusion strategies only require local communication while the network flooding is needed to apply the traditional iterative methods distributively in the mesh network. But for a large scale network, to reach an agreement among all the nodes, the consensus and diffusion strategies may take more iterations to converge to the required accuracy. Which algorithm candidates should be chosen depends on the context of the problem and the mesh network.

We also analyzed and compared the performance of the selected representative algorithms in terms of communication cost and time-to-completion. These two concerns are critical for evaluating the performance of distributed algorithms in the context of mesh networks, especially for large size problems in a large scale network due to the bandwidth, resource and time constraints in mesh networks. Besides the communication cost and time-to-completion, we think that a future research direction of distributed computing in mesh networks is the data loss tolerance: will the algorithm still approximate the optimal estimation $x^*$ well if $\alpha$-percent packets get lost in the network? Notice that, different from traditional parallel machines where data delivery is often guaranteed, in many distributed network applications, preventing data losses can either be very expensive (such as sensor networks) as it requires retransmissions, or there is a time constraint in real-time applications (such as smart grid) that makes retransmitted data useless.

\bibliography{sensorweb}



\end{document}